# Unmanned Aerial Vehicle and Optimal Relay for Extending Coverage in Post-Disaster Scenarios

Abdu Saif, Kaharudin Dimyati, Kamarul Ariffin Noordin, Nor Shahida Mohd Shah, *Member, IAENG*, Qazwan Abdullah, Mahathir Mohamad, Mahmod Abd Hakim Mohamad, and Ahmed M. Al-Saman

*Abstract*— The malfunction or interruption of wireless coverage services has been shown to increase the mortality rate during natural disasters. Wireless coverage by an unmanned aerial vehicle (UAV) provides network coverage to ground user devices during and post disaster events. The relay hops receive wireless coverage and can be forwarded to user devices that are out of coverage, allowing reliable connectivity for large-scale user devices. This work evaluates the optimal relay hops' performance to improve wireless coverage services and establish connectivity in post-disaster scenarios. The results demonstrate the UAV line of sight's understanding to select an optimal relay for improving wireless coverage services. The path loss, loss probability, and system capacity were all affected by the user device's distance and relay densities. The optimal relay hop distance and the UAV position's static are also investigated to improve coverage likelihood, which could be especially useful for UAV deployment design. It is found that the dense relays node in UAV systems enhances the capacity, coverage area, and energy efficiency by decentralized connectivity through a multi-hop device to device wireless network.

## I. Introduction

ONE of the main goals of a 5G wireless network is to support high data capacity. Despite its common usage, device to device (D2D) communication has been used in different disciplines to mean other things. D2D communication acts as a relying functionality and includes multi-hops to improve cellular downlink throughput performance. Therefore, most studies on D2D communication have focused on multi-hop relay assisted out-band D2D communications to increase overall performance. In this context, this study aims to extend the number of hops in the communication link to reduce the receiving bit error ratio efficiently, and improve the efficiency of wireless coverage services [1]–[4].

IOT-D2D communications are used under a centralized control by the cellular network to increase system capacity and energy/spectrum efficiency, and to improve wireless coverage in public safety networks [5]–[7]. In natural disaster events, coverage service to help people is essential when the wireless network is damaged or is unable to provide wireless coverage to user devices [1, 2]. The relay assisted D2D network uses a multi-hop D2D system's achievable capacity over Rayleigh fading channels. The optimal design of the D2D system and the effect of the number of relays hops on the system capacity and power efficiency are evaluated [3,4]. The role of UAVs and multi-hop D2D communications is to provide reliable connectivity in disaster situations, and to establish a communication link with user devices during out-of-coverage scenarios [5, 6].

Therefore, extending UAV coverage through the relay hop and D2D communication improves the wireless coverage service, spectrum, and energy efficiency during public safety communication [13]–[15]. The optimal relay nodes have an essential role in public safety networks by helping the UAV communicate over long distances and overcome transmission power limitations [7, 8].

This study aims to extend UAV coverage through relay hops and D2D pairs for a downlink UAV-aided wireless communication system, where D2D users coexist in an underlying manner. The UAV is able to fly and transmit wireless coverage to the relays within the disaster zone [9]. The relays forward the wireless coverage services to user devices that are out of the UAV's coverage through multi-hop D2D communications [19]–[21].

The UAV is used in public safety networks to increase the capacity and spectrum efficiency under control by the cellular

Manuscript received December 18, 2020; revised March 29, 2021. This work was supported in part by the Universiti Tun Hussein Onn Malaysia under the Multidisciplinary Research (MDR) grant vot H470 and the TIER1 Grant vot H158. Also, partly sponsored by University of Malaya under the DARE project (Grant ID: IF035A-2017 & IF035-2017).

Abdu Saif is a PhD student of Electrical Engineering Department, University of Malaya, Kuala Lumpur, Malaysia (e-mail: saif.abduh2016@gmail.com).

Kaharudin Dimyati is a Professor of Electrical Engineering Department, University of Malaya, Kuala Lumpur, Malaysia (corresponding author phone: +603-79675205; fax: +603-79561378; e-mail: kaharudin@um.edu.my).

Kamarul Ariffin Noordin is an Associate Professor of Electrical Engineering Department, University of Malaya, Kuala Lumpur, Malaysia (corresponding author e-mail: kamarul@um.edu.my).

Nor Shahida Mohd Shah is a Senior Lecturer of Faculty of Engineering Technology, Universiti Tun Hussein Onn Malaysia, Pagoh, Johor, Malaysia (corresponding author e-mail: shahida@uthm.edu.my).

Qazwan Abdullah is a PhD student of Electrical and Electronic Engineering Department, Universiti Tun Hussein Onn Malaysia, Parit Raja 86400, Johor, Malaysia (e-mail: gazwan20062015@gmail.com).

Mahathir Mohamad is a Senior Lecturer of Faculty of Applied Science and Technology, Universiti Tun Hussein Onn Malaysia, Pagoh, Johor, Malaysia (e-mail: mahathir@uthm.edu.my).

Mahmod Abd Hakim Mohamad is a Lecturer of Mechanical Engineering Department, Centre for Diploma Studies, Universiti Tun Hussein Onn Malaysia, Pagoh, Johor, Malaysia (e-mail: hakim@uthm.edu.my).

Ahmed M. Al-Saman is a Postdoctoral researcher of the Manufacturing and Civil Engineering Department, Norwegian University of Science and Technology (NTNU), Norway (e-mail: ahmedsecure99@gmail.com).

system. In any natural disaster event, communication recovery plays an essential role in public safety and saving lives [22]. Therefore, UAV has limitations for the transmission distance and power to recover disaster communication. An optimal relay is a promising approach to extend the coverage based on multi-hop D2D communication to improve wireless coverage services during disaster events [20], [23]. The UAV provides configuration to centralize beamforming to an optimal relay for reliable connectivity. It increases the signal strength at relay nodes to link with the nodes outside the coverage area. Furthermore, an optimal relay node is selected based on the residual energy and link quality in the edge of UAV coverage. An optimal relay's performance increases the coverage area by increasing the number of hops and reliably provides wireless coverage services to remote user devices.

## II. System Model

The system model is presented in Fig. 1. A scenario to help public safety networks in disaster situations is considered. The ground user devices will be distributed according to the passion point process (PPP) in the disaster area. The user device at the edge of UAV coverage will be selected as relay nodes. The relay nodes $R_i$ are able to receive wireless coverage services for the UAV and forward it to the user devices in the out of the coverage area. We assume that The optimal relay hop distance and the UAV position's static.

The multi-hop D2D communications architecture extends the UAV coverage and improves the energy and spectrum efficiency. The UAV is deployed to provide wireless coverage services to ground user devices (GUDs) in the UAV coverage area range. The UAV provides the line-of-sight coverage of the GUDs in full-duplex mode and transfers the wireless coverage signals to the cellular system out of the disaster zone area. Then, the relays $R_i$, where $(i = 1,2,....N)$ to users that will receive the wireless coverage services from the UAV and forward the wireless coverage to GUDs in out of the UAV coverage through multi-hop D2D communication.

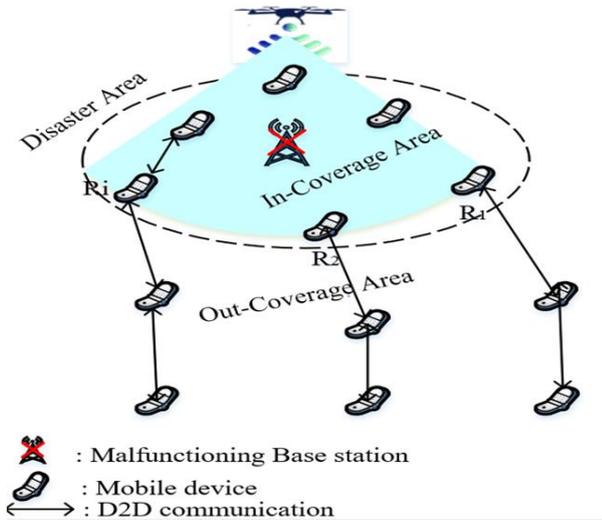

Fig. 1. System model.

This proposed model's primary goal is to verify the reliability and availability of QoS signals at ground user device receivers during natural disaster scenarios. The user devices can obtain wireless coverage services directly from the UAV. The GUDs out of the UAV coverage get indirect wireless coverage services for optimal relays directly linked to the UAV. The multi-hop D2D communication extended the coverage services for user devices far away from the UAV coverage. Therefore, the optimal relay hop is a promising technology to support the fifth-generation (5G) of wireless communications for extending the coverage area in coverage service unavailability. The optimal relays can change location dynamically to respond to an emergency and have fast reconfiguration to deliver effective communication and quicker disaster recovery.

Fig. 2 shows that user devices with the relay hops are uniformly distributed within a geographical area of 1km × 1km. The UAV represents the backhaul link for GUDs in post-disaster events. The UAV can provide full coverage to user devices through LoS in the direct communication link, and the NLoS through an optimal relay hop in the indirect communication link. Therefore, the relay is selected to be at the edge of the range in UAV coverage areas, to provide wireless coverage to the GUDs that are out of the UAV's coverage. The multi-hop D2D communication can extend the coverage and improve the wireless range in post-disaster scenarios. Furthermore, increasing the number of relay hops contributes to expanding the coverage area and system capacity, as well as improving the spectrum efficiency.

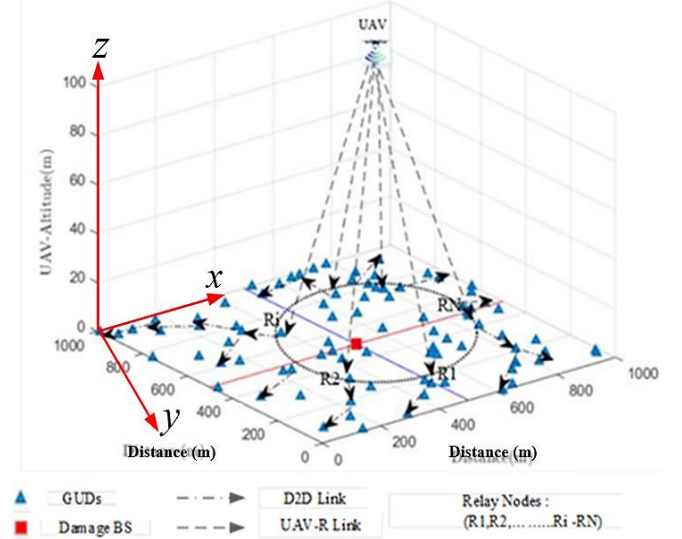

Fig. 2. Illustration location of UAV and relay hops.

### A. Probability Line of Sight (PLoS)

Line of sight (LoS) is defined as a direct communication link from UAV to ground user devices, through the air to a ground (ATG) channel. The UAV provides wireless networks with an improved probability of serving ground user devices via an LoS link, which has a lower attenuation of propagation. Besides, due to the limited transmission power and distance of the UAV to provide wireless coverage services for GUDs located in an outlying area of out UAV coverage, the optimal relays are a key to communicating with the UAV and those user devices and extend the coverage area efficiency. The UAV altitudes consider maximizing the coverage probability range and reliable connectivity in disaster situations. UAV altitude is $h$, and the radiance of coverage is $r$. Hence the UAV-GUDs distance is defined as follows:

$$d = \sqrt{r^2 + h^2} \quad (1)$$

where $d$ is the distance between the UAV and GUDs.

The possibility of an LoS and an NLoS link between the UAV and GUDs is denoted by $P_{LoS}$ and $P_{NLoS}$, respectively. This is shown as follows [10]:

$$p_{LoS}(r) = \frac{1}{1+a\,e-b\left(\frac{180}{\pi}tan^{-1}\left(\frac{h}{r}\right)-a\right)} \quad (2)$$

$$p_{NLoS}(r) = 1 - p_{LoS}(r) \quad (3)$$

where $\frac{180}{\pi}tan^{-1}(\frac{h}{r})$ is the elevation angle of the GUDs in degree, and $a$ $b$ are parameters that affect the S-curve parameters that vary according to the environment, such as urban, suburban, dense urban, and high-rise urban. Furthermore, from (2), the link is more likely to be a LoS communication link with a larger elevation angle.

*B. Path Loss Performance*

Path loss is degraded by the wireless coverage singles between the UAV-relay node links and relay-D2D communication links. Moreover, path loss propagation is a critical factor that affects the wireless channel, including the attenuation of radiated signals with distance, user device elevation angle, and the distance from the UAV. Therefore, the pathless at an optimal relay node includes Los and NLoS links with multipath fading and shadowing from the transmitted signals due to blocking and large-scale path loss obstacles. The average path loss at the relay nodes is denoted as:

$$PL(dB) = \underbrace{10\log\left(\frac{4\pi f_c d_{(UAV-r)}}{c}\right)^2}_{A} + \underbrace{\eta_{LoS} p_{LoS} + \eta_{NLoS} p_{NLoS}}_{B} \quad (4)$$

where $A$ is the free space path loss, $B$ is the average additional loss to free space loss, $f_c$ is the carrier frequency, $d\_(UAV\text{-}r)$ is the distance from the UAV to optimal relay nodes, and $c$: represents lighting speed.

*C. Optimal Relay Hop Selection*

The wireless coverage services are enhanced by selecting the optimal relay node to enable the connectivity links between UAV and GUDs in/out coverage area to reach faraway user devices in a post-disaster situation. The relay nodes are selected at the edge of coverage based on residual energy, a reliable link quality between relay nodes in the coverage area and user devices in/out of the coverage area. The UAV acts as a centralized control for all the user devices and knows every link channel in the field. Moreover, the UAV decides which user nodes can serve as the relay nodes to link functional and dysfunctional areas.

The UAV configuration adjusts the threshold for relay nodes and D2D communication for reliable connectivity in post-disaster situations for achieving efficient performance. Therefore, in performing UAV and relay hops to reduce wireless coverage services' processing ability, the call dropping problem intensifies. The call dropping occurs due to lack of one link (UAV-Ri) or (Ri-D2D) and network capacity, which varies unpredictably. The wireless coverage service will be congested, with increased communication between the user devices in/out UAV coverage and transfers the wireless coverage services to the cellular network in the core network's function area. Furthermore, the traffic congestions affect wireless coverage services' availability to keep the system running in a post-disaster situation. The LoS/NLoS channels from the (UAV-Ri) and (Ri-D2D) are assumed to be affected by the traffic transmission congested. Hence, the prominent important role here is the evaluation performance of traffic volume at peak times, which is written as follows [2]:

$$L_p = \frac{A^N k!}{N! \sum_{k=0}^{N} A^k} \quad (5)$$

where $L_p$ is the loss probability, $A$ is the offered intensity of traffic in Erlang and $N$ is the number of channels. Thus, the traffic volume's performance is evaluated based on reliable connectivity coverage distance in disaster situations. The relay nodes are assumed to be operated in time division duplex (TDD) mode. Meanwhile, one relay node can transmit simultaneously. Therefore, the relay nodes decode the received signals from UAV and forward the wireless coverage signal to the D2D communication to rejoin the out-coverage area. Furthermore, according to [3], the capacity of the D2D relay system is denoted as:

$$C_{D2D} = \frac{\lambda_d}{N} \int_0^\infty \frac{e^{-\varsigma_{dr}(\lambda_d + \gamma_{dr}\lambda_r)}}{1+v} dv \quad (6)$$

where $\lambda_d$ and $\lambda_r$ represent the D2D spatial density and relay nodes, respectively, and Table I explains $\zeta_{d_r} = C_a R_r^2 V_d^{2/\alpha}$. In the out of coverage area in Fig. 1, when the distance range of the relay changes ($R_r = \frac{R_d}{N}$), hops' rate decreases, and the number of hops $N$ increases. The power ratio of the relay node and D2D communication ($\gamma_{dr} = \left(\frac{p_r}{p_d}\right)^{2/\alpha}$) impacts the system capacity to obtain a large transmission scale for $N$-relay hops and a scalable system to establish the multi-hop communication link extending the coverage area range.

TABLE I
MATLAB SIMULATION PARAMETERS

| Symbol | Description | Values |
|---|---|---|
| $\lambda_d$ | D2D spatial density | 3.3×10⁻⁴ |
| $\lambda_r$ | Relay spatial density | (0.1- 0.5) |
| $R_d$ | D2D transmit distance | (5-50) m |
| $R_r$ | Relays transmit distance | (5- 100) m |
| $a, b, \eta_{LoS}, \eta_{NLoS}$ | Urban environment variable | 10. 6, 0.18, 1, 20 |
| $N$ | Number of hops | 10 |
| $\alpha_{D2D}$ | Path loss exponent for D2D | 3 |
| $\alpha_R$ | Path loss exponent for relay | 2 |

III. NUMERICAL RESULTS AND ANALYSIS

In this section, the simulation results are presented to analyze an optimal relay performance in extending the UAV coverage services in post-disaster scenarios. The parameters of the simulation can be shown in Table I. The proposed model is utilized to help the public safety network architecture using an optimal relay and D2D communications system to efficiently recover communication in a disaster event. The aim is to extend the UAV coverage and provide wireless coverage service to user devices between in/out coverage through multi-hop D2D communication based on the network's propagation condition in disaster scenarios. Furthermore, the optimal relay and multi-hop D2D communication are integrated with the public safety network to recover the disaster communication efficiency.

Fig. 3 shows that the path loss's performance increased when the distance between UAV and relay nodes simultaneously increased for the same level of coverage, across all carrier frequencies, due to the power consumption with ground user devices. It can also be seen that the maximum path loss was achieved at 180 m. The path loss decreased when the distance of more than 200 m increased. Thus, UAVs can increase the gain and fly over a region and operate optimally within the receiver's LoS range. Furthermore, the $f_c$ = 2.8 GHz optimal performance is due to low inference and the considerable distance it can cover. Moreover, the $f_c$ = 3.5 GHz and $f_c$ = 5.8 GHz are attributed to the path loss and interference that negatively affects the received SINR.

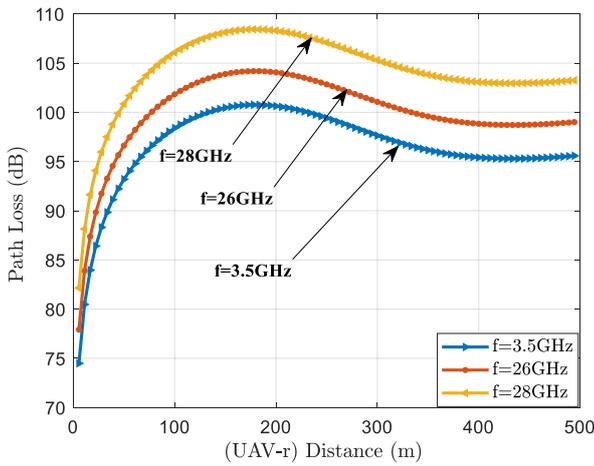

Fig. 3. path loss performance versus UAV to relay distances.

Fig. 4 shows that the performance of path loss versus the LoS probability increased with different additional path loss ($\eta_{LoS}$) that effect of PLoS, and after that decreased when the PLoS was garter than 0.2, due to different environment scenarios that affect the received signals. Fig. also shows that the path loss was affected by the PLoS in the targeted $\eta_{LoS}$. The path loss was increased from 65 dB to 110 dB, and the PLoS has risen from 0 to 1 for both targeted $\eta_{LoS}$, due to the power consumption by ground user devices. Therefore, the path loss decreased from maximum to 110 dB and 87 dB, 95 dB to 82 dB, and the PLoS increased from 0 m to 1 for the $\eta_{LoS}$ = 0.1 and 1, respectively. However, the path loss still decreased from 90 dB to around 78 dB in $\eta_{LoS}$ = 1.6 and 2.3 when the PLoS increased from 0 to 1, respectively.

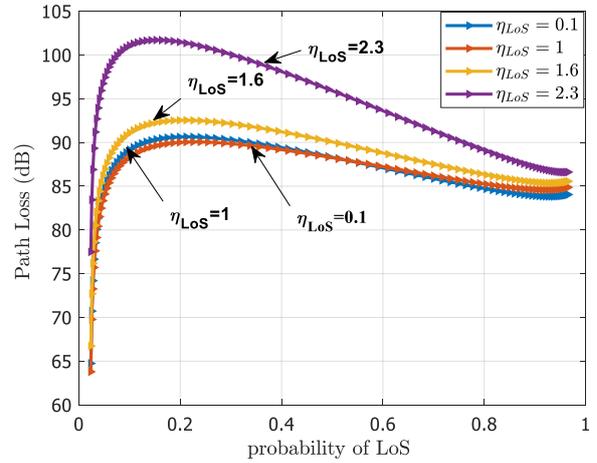

Fig. 4. The path loss performance versus probability of LoS.

Fig. 5 shows the loss probability performance versus the number of channels considered in the communication link between UAV to relay nodes and relay nodes to D2D communication. In a disaster scenario, the traffic intensity increases due to the limited network resources and the loss probability of traffic being increased. When the offered traffic intensity in Erlang $A$ = 10, the loss probability risen from 0 to 1, and several channels increased from 1 to 9. On the other hand, when $A$ = 15, 20, the loss probability grows in the same style, the number of channels increases from 1 to 10. The loss probability increased with offered traffic intensity (Erlang) per time of the call. Hence, in a disaster situation, wireless coverage availability is most important for providing reliable connectivity with user devices and keeping the network running to provide temporary coverage services.

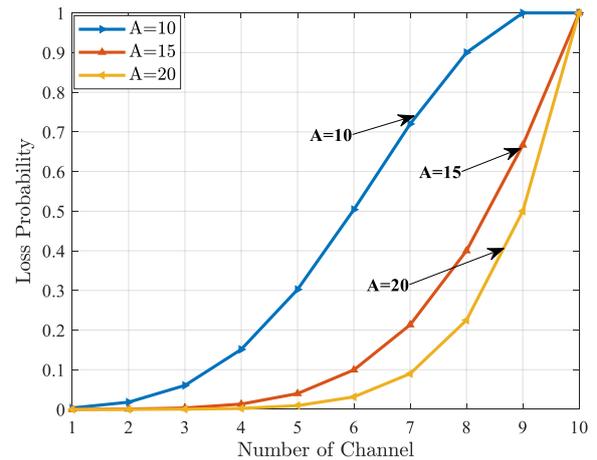

Fig. 5. Loss probability versus the number of channels.

From Fig. 6, the performance of the capacity of D2D communication increases with the number of relay hops at several relay densities, $\lambda_r$. In the figure, we observed that an increased number of hops' increases the capacity due to the increase of relay hops reducing the transmission distance on a large-scale, which achieves significant propagation gain. However, an increased number of relay hops affects the signal transmission strength. This leads to failure in the received signals at remote nodes. Hence, the UAV coverage services with the relay node densities can improve the capacity and coverage area. This enhances the energy efficiency by reducing the UAV load due to decentralized connectivity with relay nodes through the multi-hop D2D communication.

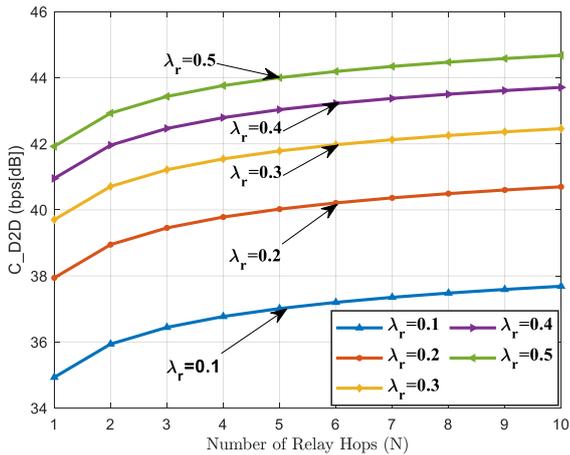

Fig. 6. Capacity of D2D versus the number of relay hops for different values of $\lambda_r$.

## IV. CONCLUSIONS

This paper improved the UAV communications model to provide emergency coverage services in disaster areas and visibility of the communications between the in-coverage and out-of-coverage areas. An optimal relay performance extended the UAV wireless coverage services, thus providing reliable connectivity in a disaster situation. The path loss, loss probability, and D2D communication capacity were analyzed based on different frequencies and relay upon densities for evaluation coverage in a disaster scenario. D2D communication capacity performance contributed to expediting the UAV network coverage area in the disaster area.